\documentclass[12pt]{spieman}  

\usepackage{amsmath,amsfonts,amssymb}
\usepackage{graphicx}
\usepackage{setspace}
\usepackage{tocloft}
\usepackage{textcomp}
\usepackage{lineno}

\usepackage{aas_macros}

\title{Demonstration of a photonic lantern focal-plane wavefront sensor: measurement of atmospheric wavefront error modes and low wind effect in the non-linear regime}

\author[a,b,c,*]{Jin Wei }
\author[a,b]{Barnaby Norris}
\author[a,b,c]{Christopher Betters}
\author[a,b,c]{Sergio Leon-Saval}

\affil[a]{Sydney Astrophotonic Instrumentation Laboratory, School of Physics, University of Sydney, NSW, Australia 2006}
\affil[b]{Sydney Institute for Astronomy, School of Physics, University of Sydney, NSW, Australia 2006}
\affil[c]{Institute of Photonics and Optical Science, School of Physics, University of Sydney, NSW, Australia 2006}

\cftpagenumbersoff{figure}
\cftpagenumbersoff{table} 
\begin{document}

\maketitle

\begin{abstract}
Here we present a laboratory analysis of the use of a 19-core photonic lantern (PL) in combination with neural network (NN) algorithms as an efficient focal plane wavefront sensor (FP-WFS) for adaptive optics (AO), measuring wavefront errors such as low wind effect (LWE), Zernike modes and Kolmogorov phase maps. The aberrated wavefronts were experimentally simulated using a Spatial Light Modulator (SLM) with combinations of different phase maps in both the linear regime (average incident RMS wavefront error (WFE) of 0.88 rad) and in the non-linear regime (average incident RMS WFE of 1.5 rad). Results were analysed using a NN to determine the transfer function of the relationship between the incident wavefront error (WFE) at the input modes at the multimode input of the PL and the intensity distribution output at the multicore fibre outputs end of the PL. The root mean square error (RMSE) of the reconstruction of petal and LWE modes were just $2.87\times10^{-2}$ rad and $2.07\times10^{-1}$ rad respectively, in the non-linear regime. The reconstruction RMSE for Zernike combinations ranged from $5.67\times10^{-2}$ rad to $8.43\times10^{-1}$ rad, depending on the number of Zernike terms and incident RMS WFE employed. These results demonstrate the promising potential of PLs as an innovative FP-WFS in conjunction with NNs.
\end{abstract}

\keywords{astrophotonics, photonics, astronomy, adaptive optics, wavefront sensors}

{\noindent \footnotesize\textbf{*}Jin Wei,  \linkable{jin.wei@sydney.edu.au} }

\begin{spacing}{2}   

\section{Introduction}
\label{sect:intro}  
The improvement of Adaptive Optics (AO) systems has been a central focus in astronomy and optical communications as a tool to correct atmospheric distortion, and more recently, to assist in the hunt for exoplanets~\cite{Davies2012}. Earth's atmosphere distorts incoming light when the light passes through regions of differing refractive index in turbulent air~\cite{Hickson2014,tyson2000introduction,tyson2012field}.  
AO's key goal is to allow terrestrial telescopes to reach imaging performance comparable to far more expensive and limited space telescopes. This is especially crucial for the coming generation of Extremely Large Telescopes (ELT), where obtaining diffraction-limited performance with their large apertures is particularly challenging.

Using AO, atmospheric turbulence's effect on the incoming light is measured in real time by a ground-based wavefront sensor (WFS), which then relays the correction to wavefront control components (e.g a deformable mirror (DM)) to correct the wavefront. Thus, the performance of the AO system is largely dependent on the accuracy of the WFS in conjunction with various reconstruction algorithms~\cite{Davies2012,Hickson2014,tyson2000introduction,tyson2012field}. 

Ideally, the wavefront would be measured in the focal plane where the science image is being formed. However, focal plane wavefront sensing must address the challenge of extracting wavefront measurements from the intensity detected at the focal plane, which is not straightforward as it lacks the crucial phase component of the image~\cite{jovanovic2018review,Barnaby2020}. As a result, a contemporary AO system generally employs a separate WFS in the pupil plane (rather than at the focal plane), called a pupil plane WFS (PP-WFS) to achieve wavefront detection.  The most widely used WFS types in current AO systems are the Shack-Hartmann WFS~\cite{Platt2001}, the pyramid WFS~\cite{Ragazzoni1996} and the curvature WFS~\cite{Roddier1988}.

There are two major disadvantages of using only a pupil-plane wavefront sensor (PP-WFS) in AO systems.  First, when using a PP- WFS, a portion of the light is redirected to the PP-WFS, usually at a different wavelength and via a different optical path, and then used to make the image. This creates two problems, chromaticity error~\cite{tyson2000introduction} and non-common path aberrations (NCPA)~\cite{Davies2012,Barnaby2020,Corrigan2016,Sauvage2007,Martinache2013,Lin:22, 10.1093/mnras/stab1634}, both which limit achievable correction of the image.  Such differences also affect the proper calibration of a WFS, which is an important step for the closed-loop performance of an AO system~\cite{tyson2012field,Lamb2014,10.1093/mnras/stab1634,Sauvage2007}. Furthermore, NCPA also manifests itself as speckle noise in high contrast extreme AO (ExAO) systems - which require very high performance to correct the wavefront to near the diffraction limit~\cite{10.1093/mnras/stab1634,Guyon2010,Olivier2018}. NCPA degrades the images by creating artifacts~\cite{Olivier2018} and is highly undesirable for ExAO systems as the speckles can resemble the signature of exoplanets. 
 
The presence of NCPA, not measurable by the PP-WFS's different optical path, is one of the major limiting factors in current ExAO systems \cite{Martinache2013,NDiaye2018,Olivier2018,Barnaby2020,10.1093/mnras/stab1634}. On the contrary, a WFS placed at the focal plane (FP-WFS) measures the wavefront directly at the same plane as the scientific camera, thus eliminating NCPA and chromacity error effect on the wavefront reconstruction.

The second major disadvantage of PP-WFSs is their limited sensitivity to certain modes, such as Low Wind Effect (LWE). LWE modes also known as island-effect modes (IE)~\cite{Milli2018, Sauvage2016, NDiaye2018,vievard2019},

and (often in the context of ELTs) petalling modes,
are described in more detail in Section \ref{subsect:LWE} of this paper. These aberrations, arising from the secondary-mirror support structure of gaps between mirror segments, present as sharp phase discontinuities in the wavefront. Thus, these effects are a major issue for AO performance, producing very strong effects in the image plane. The LWE was first reported in detail by the Spectro-Polarimetric High-Contrast Exoplanent Research (SPHERE) team~\cite{Milli2018}, reporting that they were unable to correct for this effect because the PP-WFS is only really sensitive to wavefront gradients. However a focal-plane wavefront sensor would be highly sensitive to these modes.

Various methods are being developed to achieve focal plane wavefront sensing, for example phase diversity methods \cite{Gonsalves1982,Sauvage2007,Olivier2018}, the Fast and Furious method \cite{Korkiakoski2014} and the Zernike Asymmetric Pupil Wavefront Sensor~\cite{Martinache2013}. 
However, these methods tend to rely on linear approximations -- a small incident WFE ($\ll$ 1 radian) -- which is a disadvantage as atmospheric aberrations are often in the non-linear regime, or are iterative and not well suited to rapid closed-loop correction. Another disadvantage of these FP-WFS methods is that they are not optimal for single-mode-fibre fed spectroscopy, which is desirable in advanced exoplanet characterisation.

In this paper, an all-photonic device, a photonic lantern (PL)~\cite{Leon-Saval2005,Leon-Saval2013}, is tested as a FP-WFS. The PL is placed in the focal plane, directly measuring the intensity and phase of the image. This novel fibre-based FP-WFS is also ideally suited to feed a single-mode fibre-fed spectrograph~\cite{Betters2020,Norris2022}, 
an important feature for advanced exoplanet characterisation. This configuration essentially eliminates the optical path difference between the science channel and the WFS, hence removing the NCPA.

Exploiting the full potential of PLs as a FP-WFS is an area of keen research interest.

Corrigan et al.~\cite{Corrigan2016} conducted a theoretical investigation involving computational simulations to assess the functionality of a 4-core PL as a FP-WFS. While their study primarily focused on first order aberrations, specifically tip and tilt, it yielded promising outcomes, suggesting the potential for further enhancements in performance. In a previous paper, Norris et al.~\cite{Barnaby2020} introduced the general concept of a PL wavefront sensor (PL-WFS) using a fully connected neural network (NN) to infer the input phase and intensity in the focal plane from only the output intensity of the PL.

More recently, Lin et al.~\cite{Lin:22} conducted a study into a simulated 6-mode PL and described a theoretical framework providing a promising outlook for PLs as FP-WFSs. 
A linear as well as quadratic reconstruction was described and demonstrated, and limitations identified~\cite{Lin:22}.  A separate simulation with a 6-mode mode-selective PL was also conducted and demonstrated that it was effective in characterising faint astrophysical signals, such as exoplanets~\cite{Xin2022}, as a novel nulling instrument for high contrast imaging. Norris et al.~\cite{Barnaby2020} experimentally demonstrated a 19-mode PL-WFS's ability to reconstruct wavefront error (WFE) with Zernike polynomials in a laboratory testbed. 

In this paper, we investigate the response of the 19-mode PL to the non-linear regime -- where wavefront phase error is large enough that the small-angle approximation no longer applies to the relationship between phase and wavefront sensor output-intensity. We present detailed laboratory demonstrations and characterisation of this device, being used to reconstruct various phase maps including LWE, petalling, and higher-order Zernike modes, in two different WFE regimes (linear and non-linear). Supervised NNs~\cite{lecun2015deep,SCHMIDHUBER201585} are employed for wavefront reconstruction to address the non-linear relationship between PL-WFS outputs and the incident wavefront range.

\subsection{Photonic Lanterns}
\label{subsect:PL} 
In waveguide theory, light propagates as a wave with boundary conditions through total internal reflection~\cite{Saleh:407725}, allowing a wave with a certain frequency, $f$, and a wave function,$\Psi$, to be confined in a waveguide. In astronomy, 
it was soon realised that waveguides in the form of multimode fibres (MMFs) with a large core size were able to more easily route the light from a telescope to an instrument such as a spectrograph, located some distance from the telescope's focal plane~\cite{Bland-Hawthorn:2009}. However, the field of astrophotonics demonstrated that single-mode fibres (SMFs) are important enablers that allow small, stable, replicable high-resolution spectrographs that also remove modal noise via spatial filtering.~\cite{Bland-Hawthorn:2009}. 

The PL was initially developed as a passive device to provide a low loss MMF to SMF transition (and vice versa) for astronomy \cite{Leon-Saval2005,Bland-Hawthorn:2009}. Detailed explanations of the PL are given by Birks et al.~\cite{Birks2012,Birks2015} and Leon-Saval et al.\cite{Leon-Saval2013}.
The principle of operation of the PL is based on an adiabatic transition in a tapered region of a bundle of optical fibres or waveguides~\cite{Leon-Saval2005,Leon-Saval2013}. Throughout the transition, the MM signal gradually couples to multiple SMFs (matching the number of modes) as it propagates along the transition length of the PL. The performance of a PL is also wavelength dependent due to the number of modes supported at the MM side of the PL (i.e., for the same size, the MM core will support more modes for shorter wavelengths, but the number of SM cores is fixed)~\cite{Leon-Saval2005}.

Since the excitation of each guided mode within a MMF is a function of the complex electric field at the input, the complex wavefront of the PSF is encoded within their modal coefficients\cite{Barnaby2020}. 
The PL then decouples the multiple modes into a set of single modes at the PL's SM end. 
The adiabatic transition ensures the set of decoupled SMs maintain the phase and amplitude information of the original MMs, hence the distribution and intensity of the SM outputs at the MCF end directly reflects the complex electric field input at the MMF end in the focal plane. This distribution of intensities across individual SM cores at the MCF end can be used to infer the phase and intensity information of the source, as shown at the left of the highlighted box in Fig.~\ref{fig:lantern_schem}.

\begin{figure}[pb]
 \centering
    \begin{tabular}{c c}
        \textbf{MMF end} & \textbf{MCF end} \\
        \includegraphics[width=0.4\textwidth]{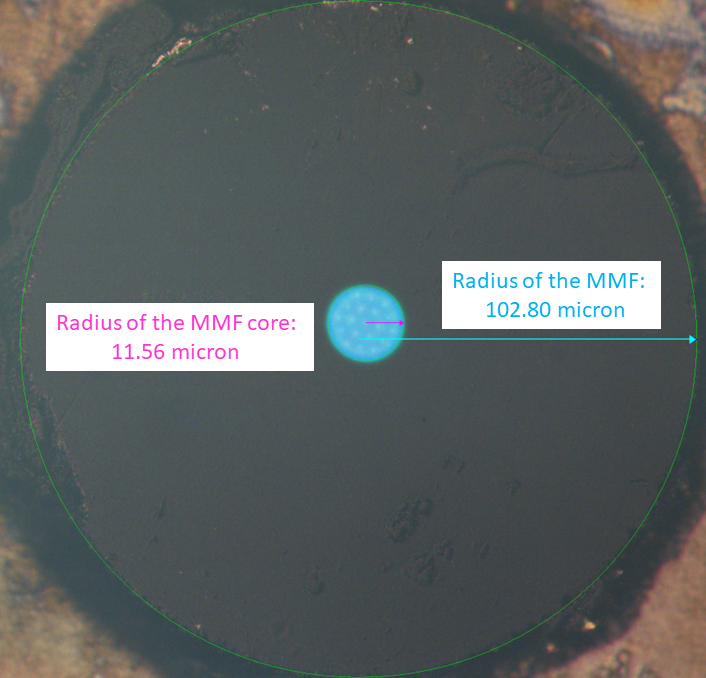} & \includegraphics[width=0.4\textwidth]{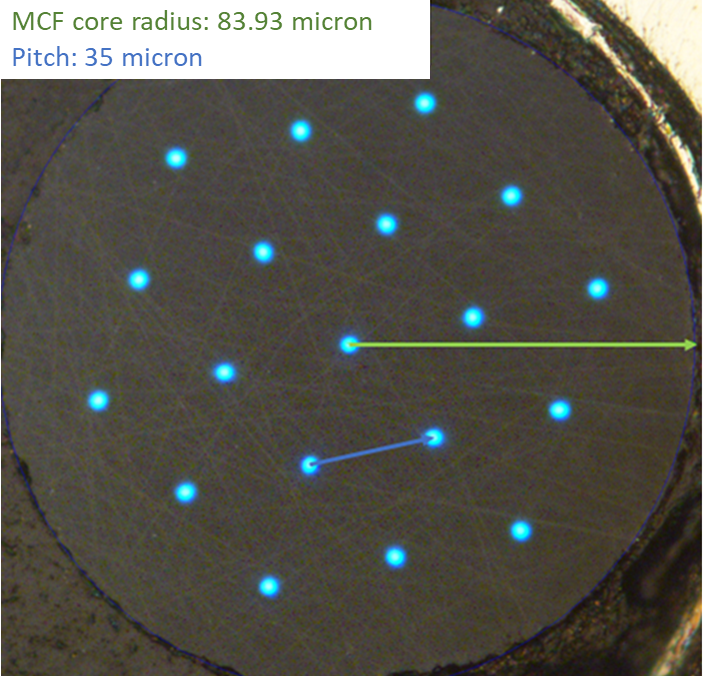}
    \end{tabular}
\caption{\textbf{Photos of the multicore-fibre-based (MCF) PL multimode fibre (MMF) and the single-mode, MCF end}: The back illuminated MMF end (left), and the illuminated MCF end showing the 19 individual SM cores (right) in a hexagonal grid with a pitch of 35 \textmu m. }
\label{fig:MMFend}
\end{figure}

The PL used in this article is a monolithic structure (seen in Fig. \ref{fig:lantern_schem}) manufactured by inserting an MCF into a capillary with a lower reflective index (RI) than the cladding of the inserted MCF~\cite{Leon-Saval2013,Birks2015,Bland-Hawthorn2016}. The composite structure (the capillary with the MCF inside) is then drawn on a tapering machine. The final product is cleaved at the taper waist to produce the MMF-SMF PLs used in this article. The MCF contains 19 SM cores in a hexagonal grid with a pitch of 35 \textmu m. Each core has a mode-field diameter of 3.78 \textmu m. The capillary and tapering process produces a MM core with an NA of 0.125 and a diameter of 23.02 \textmu m, seen in Fig.~\ref{fig:MMFend}.

\section{Experimental method}
\label{sect:exper} 
\label{subsect:method} 
\begin{figure}[pb]
\centering
\includegraphics[width=0.8\linewidth]{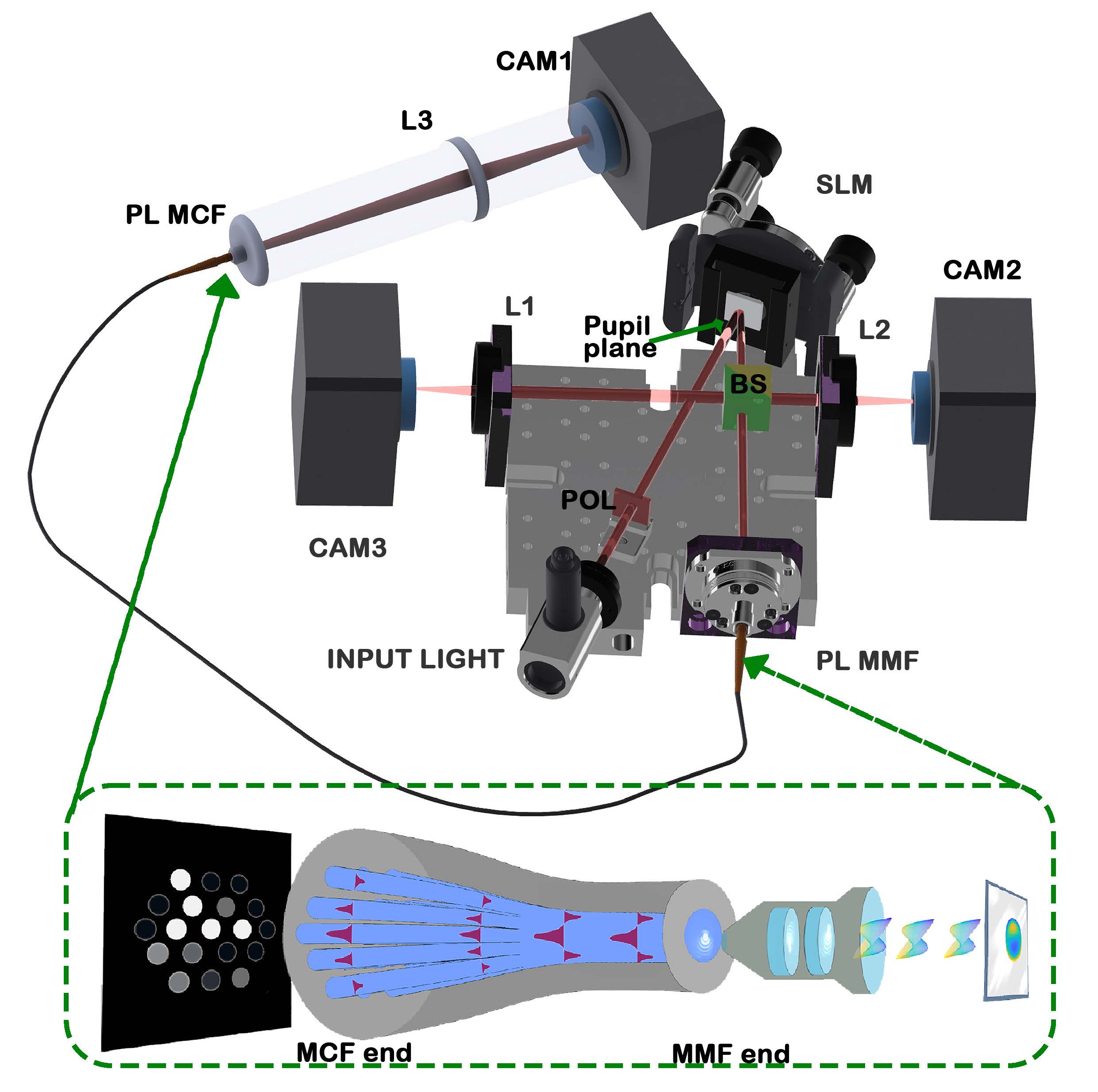}
\caption{\textbf{Schematic diagram of the PL with 19 SM outputs used in the experiments and the transition of the input wavefront from the source.} Top: schematic of the laboratory setup. Bottom: diagram of the PL with 19 SM outputs used in the experiments | From the right to left: a plane wave is reflected off an SLM (which modifies the wavefront) and is then focused by an objective lens onto the PL's MMF end, such that the size of the PSF is within the diameter of the PL's MCF end. The complex wavefront is coupled into the MMF end. The distribution of intensities in the SM cores at the output of the
MCF encode the phase and amplitude of the different modes that are coupled into the PL. Not to scale.}
\label{fig:lantern_schem}
\end{figure}

The experimental set up for testing the the PL-WFS includes a NKT SuperK Evo COMPACT supercontinuum source; a 19 core PL (as the FP-WFS); a Holoeye phase-only spatial light modulator (SLM); 3 FLIR machine vision detectors; collimation and focusing optics; opto-mechanical components; and a laptop that controls the system. The detailed setup shown in Fig. \ref{fig:lantern_schem} is placed in a temperature-controlled room with a temperature variance of +/- 0.1$^{\circ}$C peak to peak at 17.3$^{\circ}$C to reduce the effect of thermal expansion. The supercontinuum source with a 34~nm bandpass filter centred at 700~nm was used as the light source for the experiment. 

The light beam is injected into a single-mode fibre and connected to an off-axis parabolic collimator (Thorlabs RC08FC-P01) for achromatic performance. A 2 mm pupil stop is placed in the beam path to re-shape the quasi-Gaussian collimated beam into an approximated top-hat shape to eliminate the Gaussian edge; the re-shaped beam goes through the polariser (Thorlabs PBS202) and onto the SLM active surface. The total WFE of the beam then is controlled by the SLM phase-reflection output via a laptop, passes through a set of neutral-density filers to prevent over-saturation on the 3 cameras, and is then redirected by a beam splitter (non-polarising, R:I 50:50) into two beams. One beam is directed into a 200 mm focal-length lens and focused onto the first detector (FLIR Blackfly-CAM1:BFS-U3-13S2M), directly recording the PSF at the image plane produced by the SLM. The other beam is focused and injected into the MMF end of the PL with a Thorlabs Fibre Port coupler (PAF2A-A15B) (0.15 NA, 15 mm focal length). A fraction of this beam is naturally fresnel-reflected by the core of the PL MM end, allowing one to monitor the image plane at the lantern MM end via a 200 mm focal-length lens focused onto the second detector (FLIR Chameleon3-CAM2:CM3-U3-13Y3M) simultaneously as an absolute reference and for alignment. The rest of the light is coupled into the PL MM end. 
The MCF output end of the PL is then reimaged onto the third camera (FLIR Grasshopper3-CAM3:GS3-U3-32C4M) to record the SM output intensities.
All three of the cameras are set at 20.0 ms shutter speed at 0~dB gain. 10 frames were co-added for each measurement to reduce camera read-noise, and the 200~ms integration time helped average out the phase-noise produced by the SLM (arising from its use of pulse-width-modulation). Appropriate dark frames were taken for each experiment and subtracted from the measurement images.

For each modal basis to be probed, a large set of simulated wavefronts are randomly generated, with their coefficients drawn from uniform distributions of a size chosen to produce, on average, the desired wavefront error magnitude. The wavefront maps and associated coefficients are stored. Each wavefront is then applied to the SLM and the PL output intensities (as well as PSF and back-reflection camera images) are measured, with the PL output fluxes and corresponding wavefront coefficients forming the data set. 80\% of these measurements are chosen to create training data for the neural network, and the remaining 20\% withheld to be used to validate the performance of the wavefront reconstruction. This data is kept separate during the training process to avoid overfitting, which would lead to over-estimation of the reconstruction accuracy.

\section{Results and Analysis}
\subsection{Low wind effect and petalling modes measurement with the photonic lantern}
\label{subsect:LWE}

As previously mentioned, one of the advantages of a FP-WFS is the ability to detect modes such as low-wind-effect (LWE), which arise from phase discontinuities in the telescope pupil. LWE is a type of turbulence occurring inside of a telescope dome \cite{Milli2018,NDiaye2018,vievard2019}, and is attributed to radiative cooling of the telescope spiders (the structure holding the secondary mirror) caused by temperature differentials in its surroundings~\cite{Milli2018,NDiaye2018,vievard2019}. As the temperature differential modifies the refractive index of the air, the wavefront of the incoming light gets distorted due to the inhomogeneous refractive indexes of the layers of air within the telescope light path. The term `low wind effect' arises from the fact that the effect is mostly seen when windspeed is too low to break up these temperature differentials \cite{Milli2018}. Similar effects arise from the phasing-error and gaps between mirrors in segmented-mirror-telescopes (such as the ELTs), and the pupil-plane WFS's inability to constrain phase across these gaps, resulting in a per-segment phase offset. These effects can cause the PSF to break into a petal-like structure, giving rise to the term `petalling'. For clarity, in this paper we use the term `low wind effect' to refer to aberrations modelled by a tip/tilt and phase offset of pupil segments (as is conventional), and `petalling' to refer to aberrations consisting only of phase offsets between segments.

\begin{figure}[pt]
 \centering
        \includegraphics[width=0.6\linewidth]{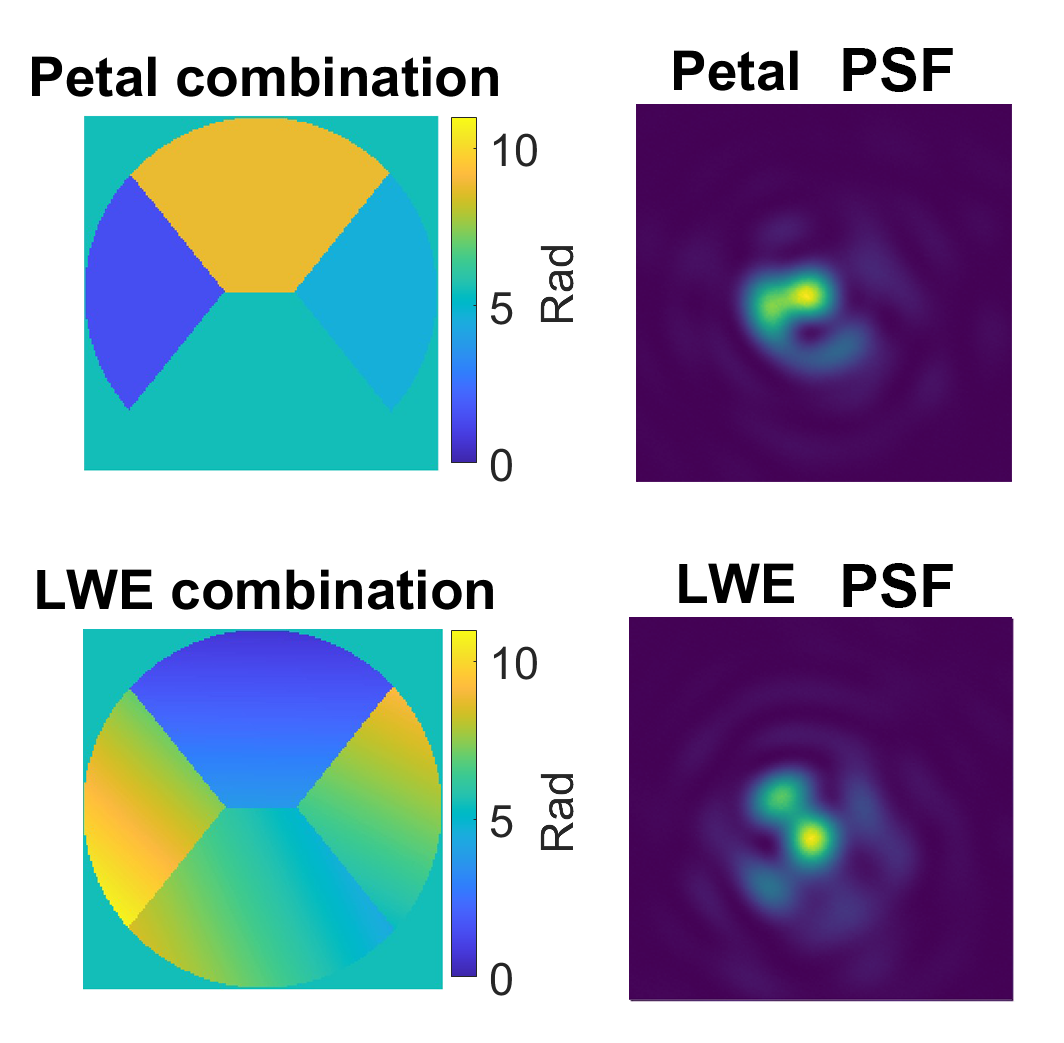}
\caption{\textbf{An example of petal-modes and LWE-modes phase maps and their corresponding PSF.} Average incident RMS WFE applied = 1.5 rad.}
\label{fig:LWE test}
\end{figure}

Here, we investigate the PL WFS's efficacy in measuring both of these aberration types. Examples of such wavefronts (drawn from the laboratory test data set) are shown in Fig. \ref{fig:LWE test}. The petal aberration (top), motivated partly by phase offsets between primary-mirror segments, is parameterised by each segment being offset by some piston term (3 parameters since global phase offset is arbitrary). The LWE aberration (bottom) adds tip and tilt components to each segment, as is conventionally used to simulate this effect\cite{vievard2019}.

First petal-modes were investigated, and 80,000 petal phase maps containing a total incident root-mean-squared wavefront error (RMS WFE) ranging from 0 to 3.1 rad (1.5 rad on average) were randomly generated. This WFE range was constrained by the maximum peak-to-valley phase modulation achievable by the SLM used, which was $\sim$11~rad. These were applied to the SLM and the resulting PSFs injected into the PL to produce a data set as described in Section \ref{subsect:method}.

This data was subsequently split into a training set and test set (comprised of 80\% and 20\% of the data respectively). A fully-connected neural network was then trained (using only the training data) to predict the wavefront coefficients for a given set of PL output fluxes. The hyperparameters dropout rate, learning rate and the number and size of hidden layers were tuned to minimise overfitting and produce robust predictions, with the resulting hyperparameters given in Table~\ref{tab:NN petal construct}. For all NNs described here, a ReLU activation function and the Adam optimizer were used, with a learning rate of $1\times10^{-4}$. 

\begin{table*}[pb]
\centering
\begin{tabular}
{ p{0.07\linewidth} p{0.13\linewidth} p{0.13\linewidth} p{0.22\linewidth} p{0.07\linewidth} p{0.07\linewidth} p{0.13\linewidth} }
\hline
  Zernike Terms & Incident RMS WFE (rad) & Incident Peak-Valley (rad) & Number of layers /neurons & Epochs & Dropout rate & Total\newline reconstruction RMSE (rad) \\
         \hline 
         Petal modes & 1.50 & 4.66 &1000-100-3 & 100 & 0.05 &  $2.87\times10^{-2}$\\
         \hline
         LWE & 1.50& 6.94 &12000-2000-100-11  &  100 & 0.15 &  $2.07\times 10^{-1}$\\

    \hline
\end{tabular}

\caption{\textbf{Wavefront reconstruction accuracy results for petal and LWE modes.}  Reconstruction error (for incident wavefront error of 1.5 rad RMS) is given as root-mean-square error (RMSE) in radians. Hyperparameters for the utilised NN are also given.}
\label{tab:NN petal construct}
\end{table*}

\begin{figure}[pt]
    \centering
    \includegraphics[width=0.8\linewidth]{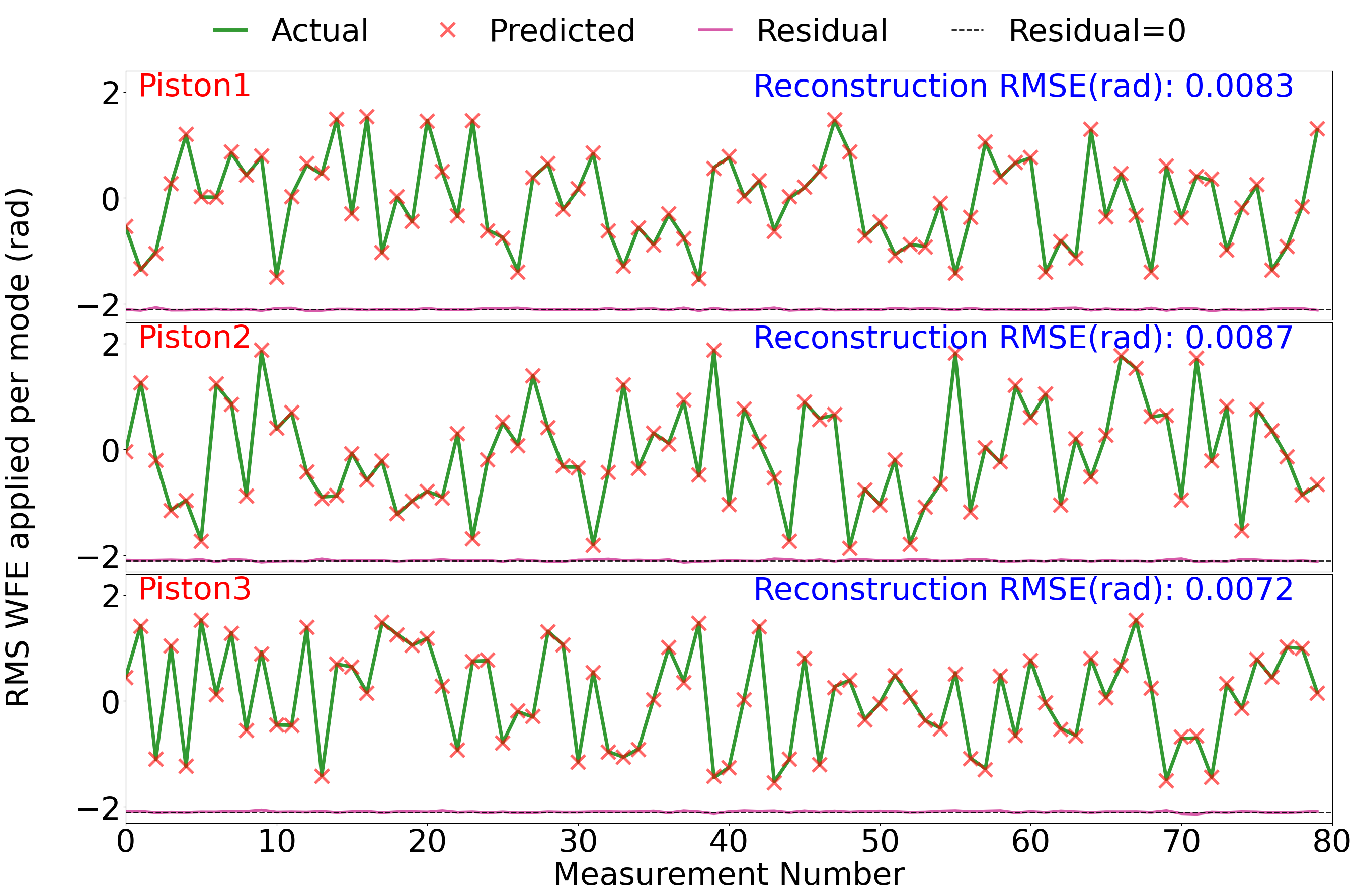}
    \caption{\textbf{Comparison between the actual coefficients and predicted coefficients by the NN for petal modes with average input RMS WFE of 1.5 rad}: the residuals of each modes are the magenta dotted line (almost indistinguishable from the axis of "Residual = 0"). }
    \label{fig:Peddle predit}
\end{figure}
\begin{figure}[pb]
\centering
    \includegraphics[width=0.8\linewidth]{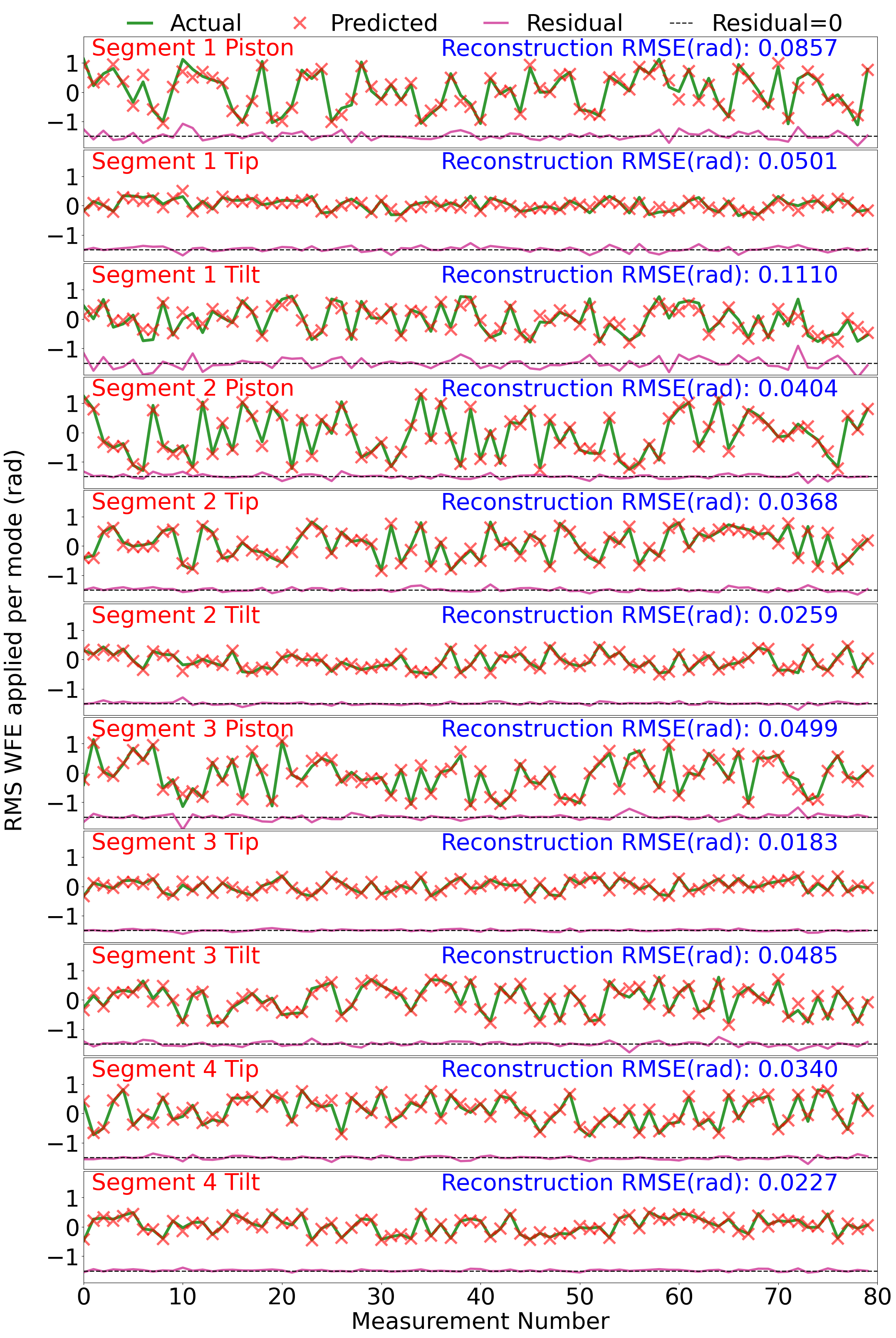}
    \caption{\textbf{Comparison between the actual coefficients and predicted coefficients by the NN for LWE modes with average input RMS WFE of 1.5 rad}: the residuals of each modes are marked with the magenta dotted line (almost indistinguishable from the axis expect the first 3 modes).}
    \label{fig:LWE predit}
\end{figure}
This trained neural network is then used to predict the wavefront coefficients for previously unseen test data, and the root-mean-squared-error (RMSE) between the true and predicted wavefronts is calculated, yielding an RMSE of $2.87\times10^{-2}$ rad, which translates into a residual of 3.20 nm (at 700 nm wavelength). A graph showing the actual versus predicted coefficients for the three petal modes is given in Fig. \ref{fig:Peddle predit}, and a correlation plot is given in Fig. \ref{fig:zernike correlation}. 
This is a very accurate reconstruction, demonstrating the PL's intrinsic sensitivity to petal type modes (aided by the fact that the 19 output fluxes are being used to constrain only 3 free parameters).

An expanded LWE modal basis, containing 11 parameters, was then tested. 
 LWE as an aberrated wavefront can be described analogously as combinations of piston, tip and tilt modes \cite{Milli2018}, motivated by the phase discontinuities across the telescopes' secondary mirror supports. The 11 piston, tip and tilt modes were randomly generated, each drawn from a uniform distribution such that the resulting WFE ranged from 0.5-3.2 rad with an average of 1.5 rad
 (with one segment having no position term since global phase is undefined).
 The NN with hyperparameters as per Tab. \ref{tab:NN petal construct} was trained in the same way as the previous example, and resulted a total reconstruction RMSE of $2.07\times10^{-1}$ rad or 23.06 nm (at 700 nm wavelength). Fig. \ref{fig:Peddle predit} shows one example of the reconstructed phase map from the predicted coefficients obtained from the above method; the reconstruction correlation plot is given in Fig. \ref{fig:zernike correlation}, demonstrating the efficacy of the PL WFS in sensing these modes in the non-linear regime ($\sim$10 rad P-V wavefront error).

\begin{figure}[pt]
    
    \centering
    \includegraphics[width=0.7\linewidth]{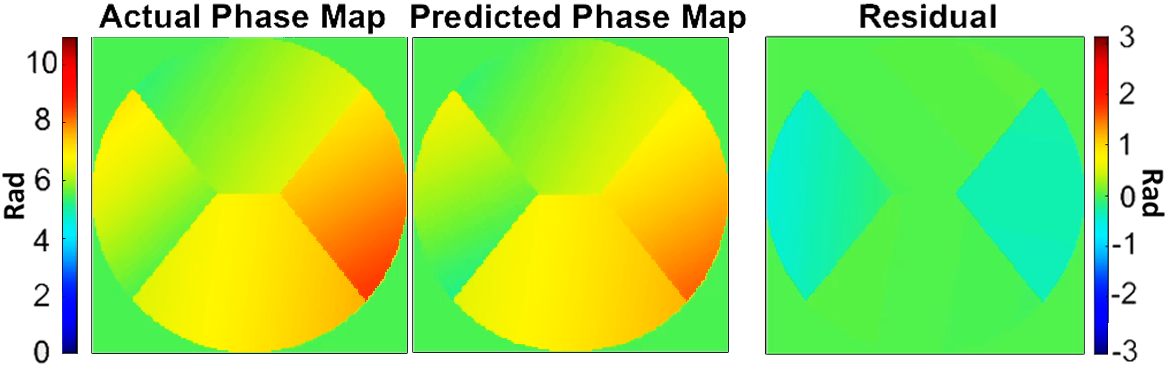}
    \caption{\textbf{A visual comparison between the actual and predicted phase map for LWE.} Note the scale for the residual is smaller than the actual and predicted phase maps in order to make visible the reconstruction error. }
    \label{fig:LWE residual}
\end{figure}

\subsection{Zernike modes measurements with the photonic lantern}
\label{subsect:zernike} 
Zernike polynomials are widely used as a mode basis to describe low-order optical aberrations\cite{tyson2000introduction}. Here, a set of 9, 14 and 19 Zernike terms were used to generate phase maps on the SLM for two incident RMS WFE regimes, for a total of 6 different measurement sets. The collection method for this data was the same as the previous petal and LWE experiments.

NN hyperparameters were optimised for each case, with their details given in  Tab.~\ref{tab:NN construct}. 

\begin{figure}[pht]
    \centering
    \includegraphics[width=0.5\linewidth]{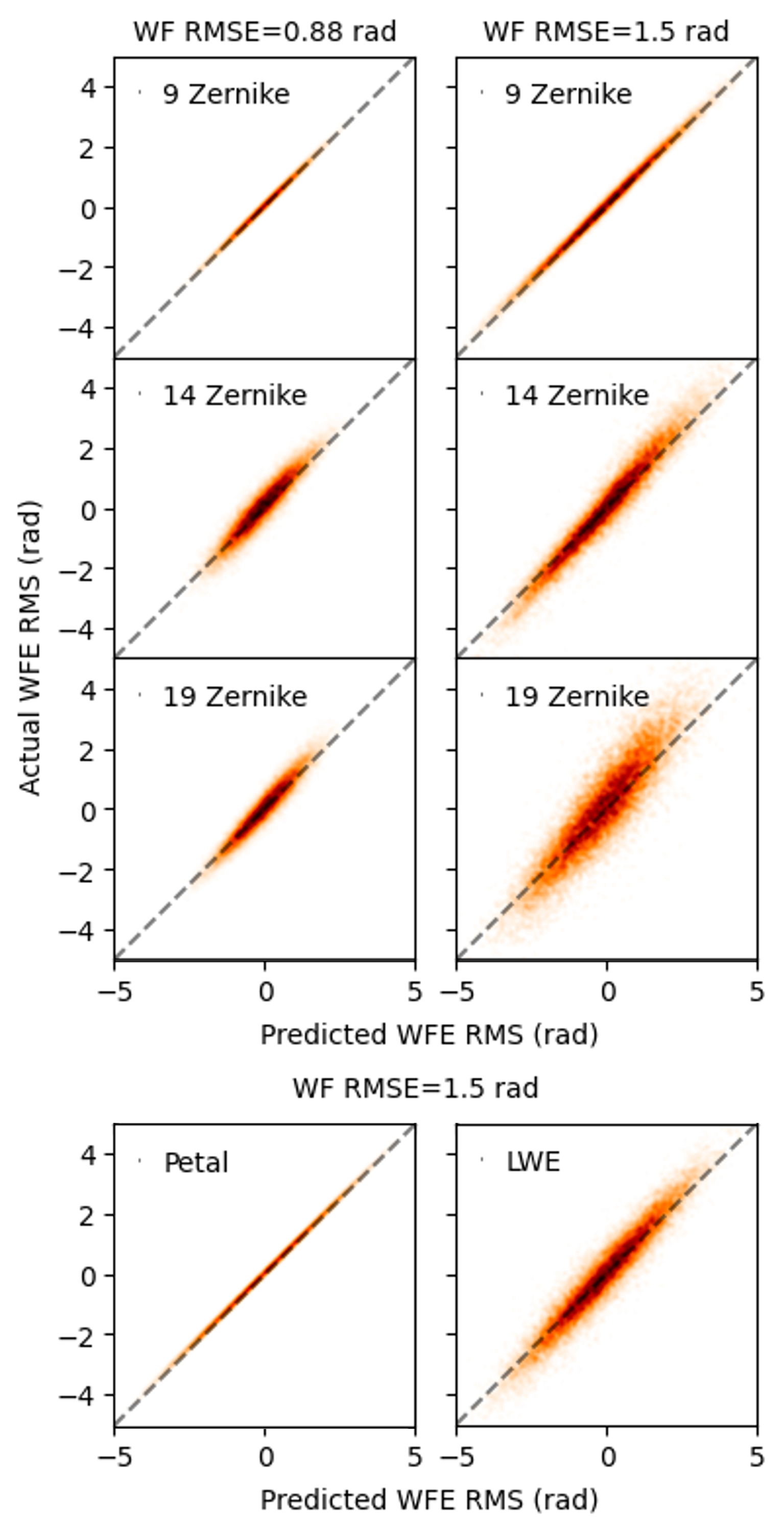}
    \caption{\textbf{Correlation histograms for all experiments, showing the reconstruction accuracy.} The predicted coefficients closely match the actual coefficient values for the 9 Zernike phase map at lower RMS WFE. As the number of Zernike term and RMS WFE increases, the scatter becomes larger. The dotted line is the theoretical perfect fit.}
    \label{fig:zernike correlation}
\end{figure}

The coefficients of the Zernike modes were randomly generated from a uniform distribution to form the final phase map projected onto the SLM. The Zernfun function~\cite{Paul2021} was used to generate the Zernike phase maps from the modal coefficients.
In the first set, the maximum coefficient value was set such that the combined Zernike modes yielded an incident RMS WFE of 0.88 radian on average. To probe reconstruction outside the linear-approximation regime another set of the same number of Zernike modes (9, 14, and 19) with larger coefficients was similarly generated, which yielded phase maps with a total RMS WFE of 1.5 radian on average.

\begin{figure}[pb]
    \centering
    \includegraphics[width=0.7\linewidth]{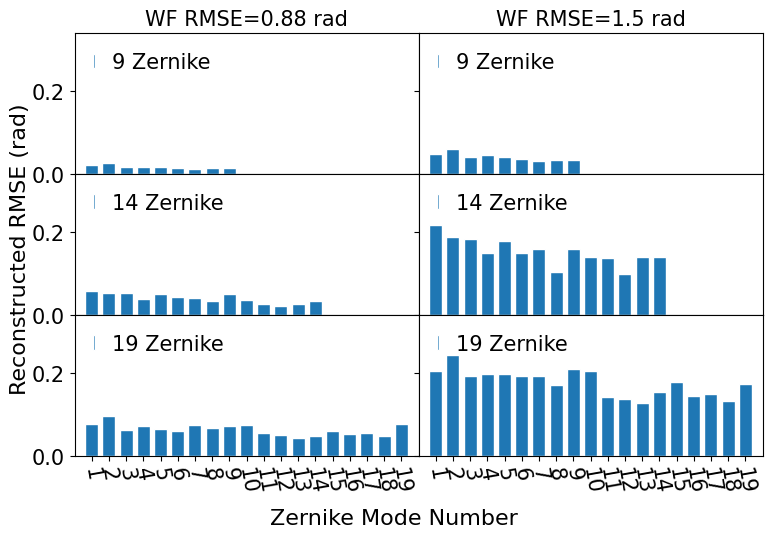}
    \caption{\textbf{Reconstruction RMSE per mode for the Zernike-basis tests, for three different numbers of terms and for both the low and high incident WFE regimes.}}
    \label{fig:zernike rmse}
\end{figure}

\begin{figure}[pt]
    \centering
    \includegraphics[width=0.8\linewidth]{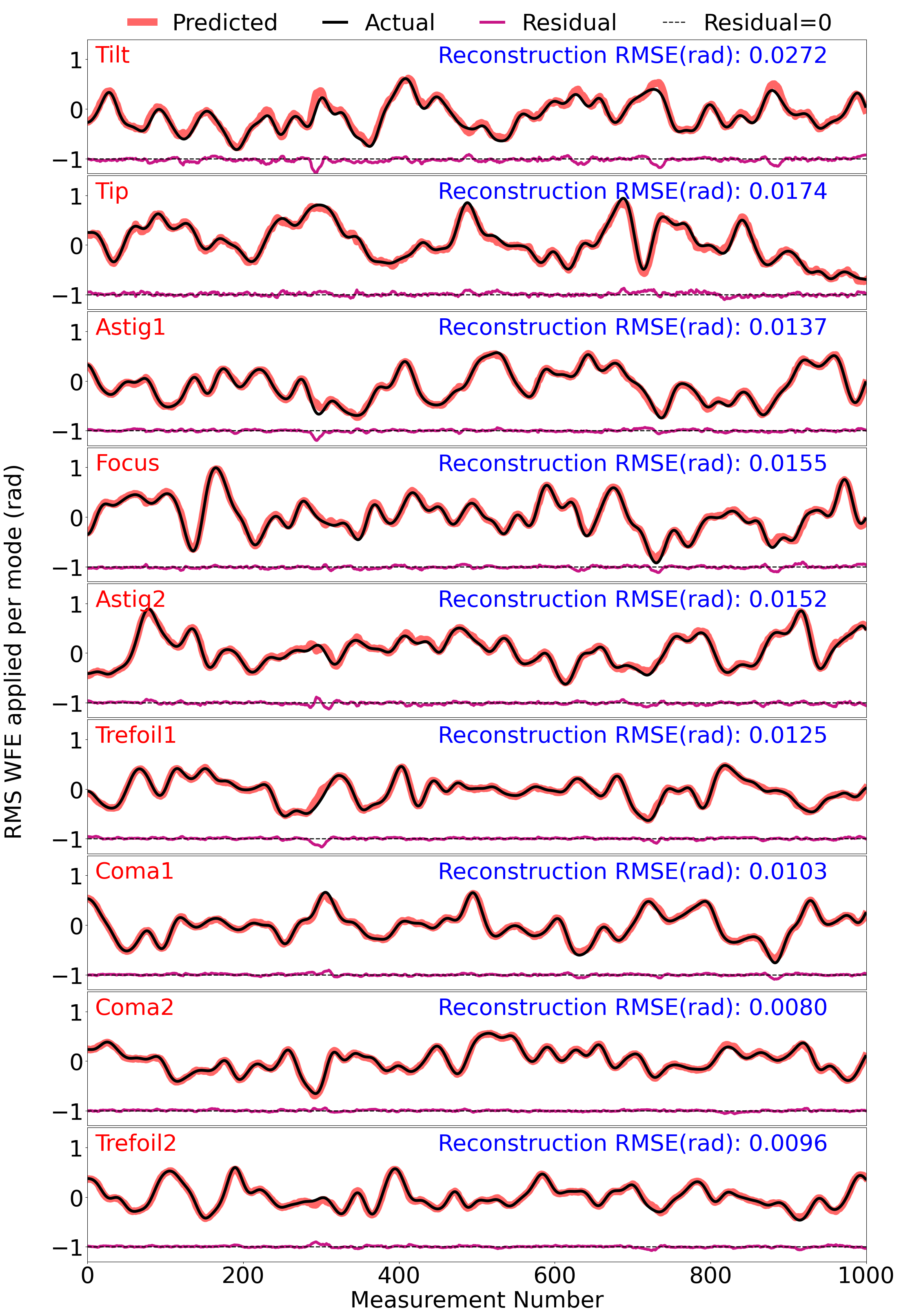}
    \caption{\textbf{A time-series line plot for a 9 Zernike term reconstruction test, for the low incident WFE regime. The applied wavefront coefficients for the test data vary continuously with time.}}
    \label{fig:9 zernike time series}
\end{figure}

\begin{figure}[pb]
    \centering
    \includegraphics[width=0.8\linewidth]{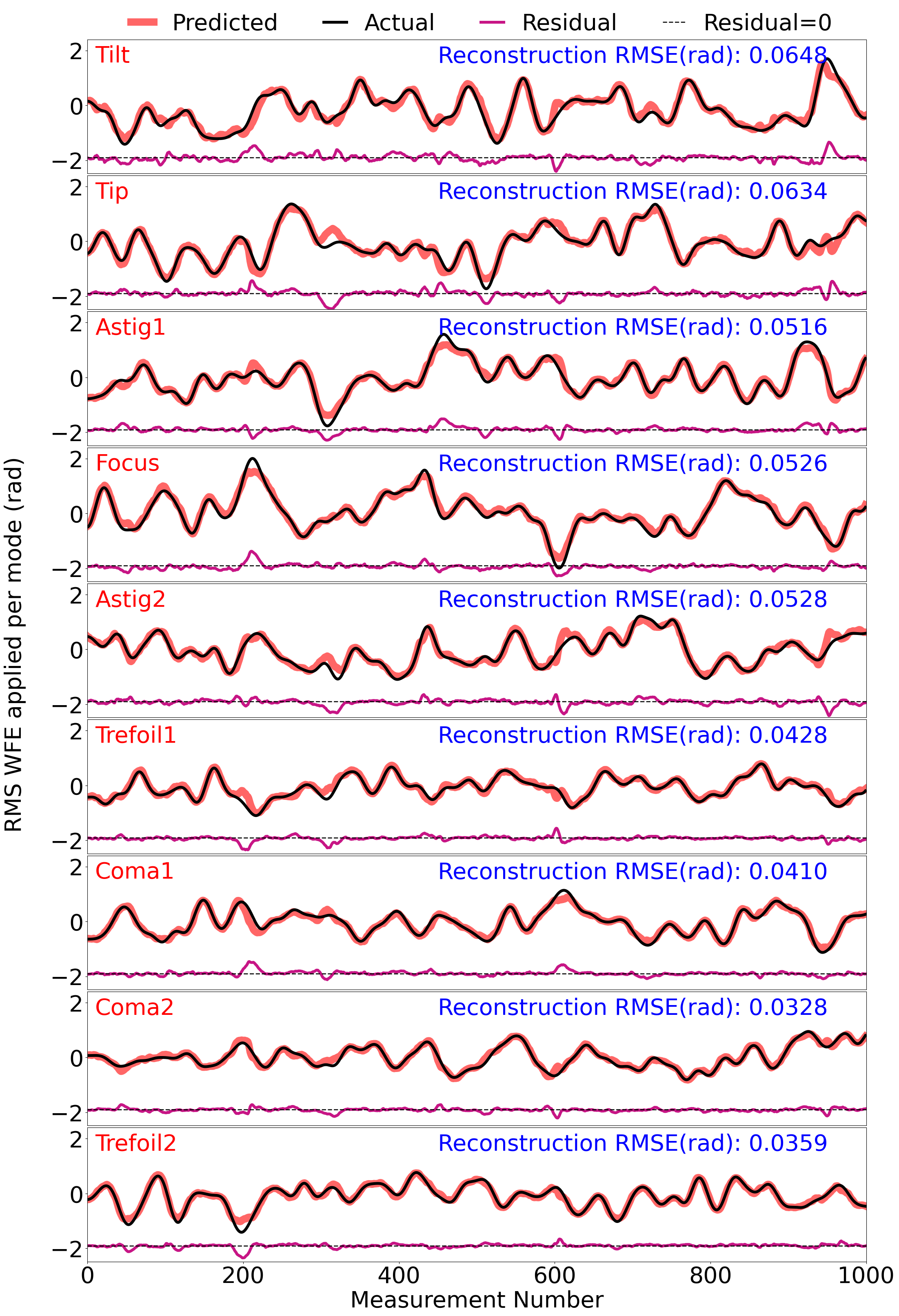}
    \caption{\textbf{A time-series line plot for a 9 Zernike term reconstruction test, for the high incident WFE regime. The applied wavefront coefficients for the test data vary continuously with time. When compared to the low incident WFE test, departures of the predictions from true values are more visible, occurring mostly in regions where the magnitude of the WF coefficient is large.}}
    \label{fig:9 L zernike time series}
\end{figure}

For each combination of incident WFE and number of Zernike terms, 80,000 measurements were taken. These measurements were then divided into training and testing datasets to facilitate the training and evaluation of a NN as in previous experiments. The resulting reconstruction accuracies are given in Tab.~\ref{tab:NN construct}, and shown in the correlation plots in Fig. \ref{fig:zernike correlation}.  
\begin{table*}[phb]
\centering
\begin{tabular}
{ p{0.07\linewidth} p{0.13\linewidth} p{0.13\linewidth} p{0.22\linewidth} p{0.07\linewidth} p{0.07\linewidth} p{0.13\linewidth} }
\hline

  Zernike Terms & Incident RMS WFE (rad) & Incident Peak-Valley (rad) & Number of layers / neurons & Epochs & Dropout rate & Total\newline reconstruction RMSE (rad) \\
         \hline 
         9 & 0.88 &  5.06 &2000-1100-100-9 & 100 & 0.004 & $5.67\times10^{-2}$\\
         \hline
         9 & 1.50& 8.80 & 2000-2000-200-9  &  100 & 0.05 & $1.53\times10^{-1}$\\
         \hline
         14 & 0.88 &5.68 & 2000-2000-2000-14 &  100  & 0.22 & $1.86\times10^{-1}$\\
         \hline
         14 & 1.50 & 10.04 & 2000-2000-2000-14 &  300  & 0.40 & $5.86\times10^{-1}$ \\
         \hline
         19 & 0.88 & 5.98 & 2000-2000-2000-19 &  200  & 0.40 & $3.43\times10^{-1}$\\
         \hline
         19 & 1.50 & 10.61 &2000-2000-2000-19 &  200  & 0.40 & $8.43\times10^{-1}$\\
    \hline
\end{tabular}

\caption{\textbf{Reconstruction accuracy, in terms of RMSE (rad), for each of the 6 cases of Zernike terms and their corresponding NN hyperparameters}}
\label{tab:NN construct}
\end{table*}
In addition to the randomly chosen coefficients, additional test data sets (9 Zernike modes with the two incident WFE regimes) were acquired where the Zernike coefficients varied continuously over time (produced by filtering noise by a Gaussian kernel), analogous to the continuously varying wavefront encountered in actual seeing. This test data was used to reconstruct the wavefront using the already trained NNs, with the results presented in  Fig.~\ref{fig:9 zernike time series} and~\ref{fig:9 L zernike time series}. 

It can be seen that for the low incident WFE case the predicted coefficients very closely match the true values, with each individual coefficient's RMS reconstruction error being on average 0.014~rad (1.6~nm at $\lambda=700$~nm). For the high incident WFE regime regions of poorer reconstruction are visible, and largely occur when the wavefront coefficient is large ($\gtrapprox$ 1 rad), with each coefficient having on average an RMS reconstruction error of 0.049 rad (5.5~nm). This suggests that there may be further improvement possible in NN architecture design to better account for the non-linearity as applied RMS WFE increases.

The results from Tab.~\ref{tab:NN construct} and Fig. \ref{fig:zernike correlation} show that the PL WFS and NN perform well in both WFE regimes, but have reduced accuracy in the high WFE regime. This is consistent with expectations, as the relationship between wavefront phase and PL output is quite non-linear in the latter case. While the NN predicts well in this non-linear regime, further optimizing NN architectures or hyperparameters could enhance its performance.
It is also seen that reconstruction accuracy decreases as the number of modes used increases, as would be expected as the same quantity of data (with the same signal-to-noise ratio) is being used to constrain more degrees of freedom. It should also be noted that the modal basis of the PL is not necessarily isomorphic with the Zernike basis of the applied wavefront error, so one would expect an accurate representation of $n$ Zernike terms to require $>n$ PL output fibres.

Fig.~\ref{fig:zernike rmse} shows the contribution to reconstruction RMSE (rad) from each mode, for each of the Zernike-basis experiments. While the distribution of error is roughly even, some trends can be discerned. In many cases (especially evident in the 9 and 19 Zernike plots) the first two modes, corresponding to tip and tilt, have higher errors. It is suspected this is due to drifts in alignment and vibration in the laboratory setup, introducing tip/tilt variation in the training data which is not reflected in training labels. Also there appears to be a general trend where the higher order terms have lower errors. This may be because the effect of beam tip/tilt on the powered optics in the setup tends to induce lower order aberrations rather than higher order ones.

Lastly, a qualitative experiment was performed using 19 Zernike terms to reconstruct phase maps of Kolmogorov turbulence, representing astronomical seeing. The phase maps are translated across the pupil over time, corresponding to wind speed (the `frozen flow' model of seeing). Kolmogorov phase maps were constructed using HCIPy \cite{por2018hcipy} (with an average incident RMS WFE of $\sim0.9$ rad) and displayed on the SLM, and data taken in the same way as the other experiments. The wavefront was then inferred from these new PL fluxes using the existing NN which had been trained on the 19 Zernike data (described above). 
The PL WFS would be expected to reconstruct a low-order approximation of the actual Kolmogorov phase map, since that contains much higher spatial frequencies than those sampled by the PL. Thus to provide a useful visual reference, each phase map was decomposed into a Zernike basis representation (using OpticsPy~\cite{Fan2019OpticsPy}) while discarding all components higher than the 19th Zernike mode. This essentially gives the best-case representation using the available basis. 
\begin{figure}[pt]
    \centering
    \includegraphics[width=0.8\linewidth]{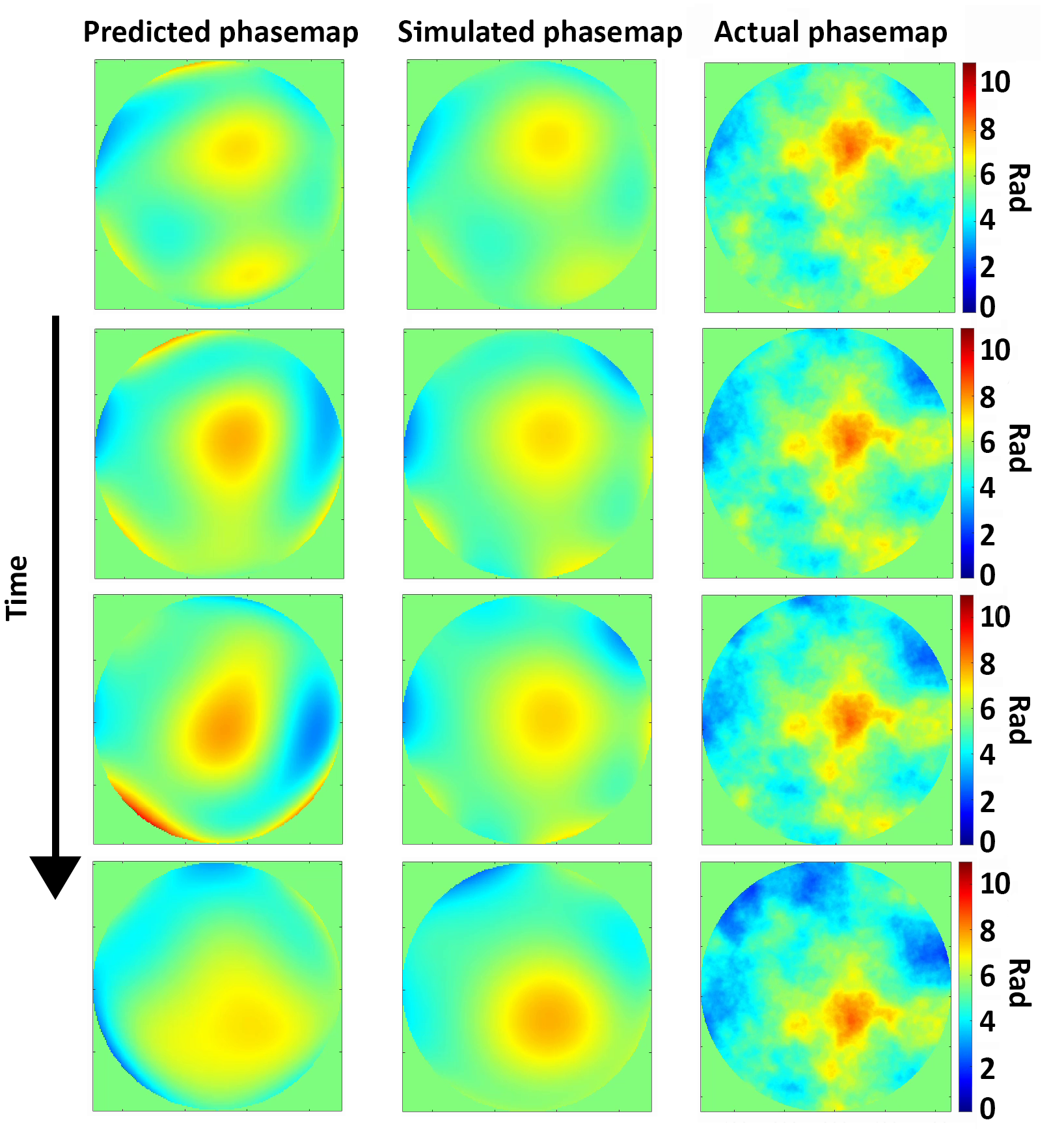}
    \caption{\textbf{Reconstruction of of a moving Kolmogorov turbulence phase screen with the PL WFS.} Right: the actual applied wavefront. Centre: the applied wavefront represented by only the first 19 Zernike modes, essentially the `best-case' reconstruction achievable with this basis. Left: the wavefront reconstructed by the PL WFS. Future PLs with more modes/outputs will allow higher spatial frequencies to be reconstructed. }
    \label{fig:tubulance}
\end{figure}
An example of four frames from this time sequence is given in Fig. ~\ref{fig:tubulance}, including the true phase map applied, the true phase map composed only of the first 19 Zernike terms (a.k.a. `simulated' phase map), and the phase map reconstructed by the PL WFS.

\section{Conclusion}
\label{section:conclusion}  

In this study, we illustrate the promising capabilities of PLs paired with NNs to act as effective FP-WFSs for existing and future AO systems. FP-WFSs are crucial for AO systems as they can correct NCPAs errors, and we demonstrate the photonic lantern wavefront sensor (PL-WFS)'s ability to measure these low-order aberrations even with high wavefront error, outside the linear range. 
This is particularly advantageous for addressing modes, like the LWE, that are problematic for extreme AO systems and are in the null space of common pupil-plane wavefront sensors. A PL-WFS is also ideal for cases where efficient injection into a single mode fibre is desired (such as precision spectroscopy) and provides optimal detector noise properties, using only 1 pixel per spatial mode. It also lends itself well to future fully-photonic adaptive optics systems, where the uncorrected PSF is injected into the PL and the phase and amplitude of its outputs manipulated and recombined with active photonic circuitry. 

We demonstrate here that the PL can successfully predict up to 19 Zernike terms with an average incident RMS WFE as high as 1.5 rad, resulting in a total reconstruction RMS error between \(5.67 \times 10^{-2}\) rad (6.3~nm at $\lambda=700$~nm) and \(8.43 \times 10^{-1}\) rad (94~nm), depending on the number of Zernike terms and the incident RMS WFE. Reconstruction of problematic LWE modes in the non-linear regime (average RMS WFE=1.5 rad, or 7~rad P-V) was also demonstrated, with a reconstruction RMSE of only \(2.07 \times 10^{-1}\) rad (23.1~nm).

Scaling the PL to a larger number of modes, in order to reconstruct higher order wavefront error, is a clear next step. Spectrally dispersing the PL's fiber outputs is optically trivial, and using wavelength-dependent information in AO correction (e.g. for phase unwrapping) should also be tested. With sufficient mode count, complete spectro-spatial image reconstruction is within reach. These and other developments of the PL-WFS concept stand to offer substantial benefits across diverse research fields, ranging from astronomy to optical communications.

\section*{Data availability} 
Data underlying the results presented in this paper are not publicly available at this time but may be obtained from the authors upon reasonable request.

\section*{Acknowledgements} 
B.Norris is the recipient of an Australian Research Council Discovery Early Career Award (DE210100953) funded by the Australian Government.

\smallskip

\bibliographystyle{spiejour}
\bibliography{bibliography}

\section*{Biography}
 \vspace{2ex}\noindent\textbf{Jin Wei} is a MPhil student currently in the process of completing her degree in University of Sydney. She is the co-author for previously published work on photonic lanterns as wavefront sensors.

\end{spacing}
\end{document}